\documentclass[preprint,showpacs,aps]{revtex4}

\usepackage{epsfig}

\begin{document}

\title{\large \bf Non-equilibrium critical dynamics with domain wall and surface}

\author{\bf N. J. Zhou$^{1,2}$ and B. Zheng$^{1,}$\footnote{corresponding author: zheng@zimp.zju.edu.cn}}

\affiliation{$^1$ Zhejiang University, Zhejiang Institute of Modern
                  Physics, Hangzhou 310027, P.R. China\\
             $^2$ Department of Physics, National Central University,
                       Chungli, Taiwan 320}

\begin{abstract}
With Monte Carlo simulations, we investigate the relaxation
dynamics with a domain wall for magnetic systems at the critical
temperature. The dynamic scaling behavior is carefully analyzed,
and a dynamic roughening process is observed. For comparison,
similar analysis is applied to the relaxation dynamics with a free
or disordered surface.
\end{abstract}


\maketitle

\section{Introduction}

In the past years, much effort of physicists has been devoted to
the understanding of non-equilibrium dynamic processes. Phase
ordering dynamics, spin glass dynamics, structural glass dynamics
and interface growth etc are important examples. Since the pioneer
work by Janssen et al \cite{jan89}, the universal dynamic scaling
form in critical dynamics has been explored up to the {\it
macroscopic} short-time regime
\cite{jan89,hus89,hum91,sta92,ito93,li94,li95,luo98,zhe98,zhe99},
when the system is still far from equilibrium. Although the
spatial correlation length is still short in the beginning of the
time evolution, the short-time dynamic scaling form is induced by
the divergent correlating time around a continuous phase
transition. Based on the short-time dynamic scaling form, new
methods for the determination of both dynamic and static critical
exponents as well as the critical temperature have been developed
\cite{blu92,sta92,ito93,li95,luo98,zhe98,zhe99}. Since the
measurements are carried out in the short-time regime, one does
not suffer from critical slowing down.

In understanding the dynamic scaling form far from equilibrium, we
should keep in mind that it holds after a time scale $t_{mic}$,
which is sufficiently long in the microscopic sense, but still
short in the macroscopic sense. More importantly, the {\it
macroscopic} initial condition should be taken into account in the
dynamic scaling form \cite{jan89,zhe98,fed06}. For the dynamic
relaxation starting from an ordered state, i.e., a state with an
initial magnetization $m_0=1$, for example, the magnetization
decays by a power law \cite{luo98,zhe98,fed06}. If $m_0$ is
smaller but close to $1$, there emerge corrections to scaling. For
the dynamic relaxation starting from a random state, i.e., a state
{\it without} spatial correlations and with a small $m_0$,
however, the magnetization does not decay, and rather shows {\it
an initial increase} in the macroscopic short-time regime. An
independent critical exponent $x_0$ must be introduced to describe
the scaling dimension of the initial magnetization
\cite{jan89,li94,zhe98,zhe99}. If $m_0=0$, the magnetization
naturally remains zero during the dynamic evolution, but $x_0$ is
still needed to describe the auto-correlation function etc. This
critical exponent also explains the power-law decay of the
remanent magnetization in spin glasses \cite{hus89,rie93,luo99}.
On the other hand, the short-time dynamic scaling form is {\it
universal}, in the sense that it does {\it not} depend on the
microscopic details of the dynamic system, such as the lattice
types, interactions, and updating schemes etc. Up to now, the
dynamic relaxation with the {\it ordered} and {\it random} initial
states has been systematically investigated.

Recent progress in the non-equilibrium critical dynamics and its
applications includes, for example, theoretical calculations and
numerical simulations of the XY models and Josephson junction arrays
\cite{zhe03a,oze03b,gra05,nie06}, magnets with quenched disorder
\cite{yin04,yin05,oze05,che05,yin06}, and various critical systems
\cite{alb02,gra04,lan05,ara06}. Dynamic reweighting methods have
been proposed \cite{lee05,yin05}, and the dynamic approach to the
weak first-order phase transitions is also attractive
\cite{sch00,alb01a,sar03,yin05}. Recently, the ageing phenomenon
around a continuous phase transition has been also intensively
studied \cite{god02,hen04,pic04,sch05,ple05b,cal05,lei07}. In this
case, the dynamic scaling form for the ageing phenomenon is induced
by the long-range time correlation, different from that induced by
meta-stable states in glassy systems below the transition
temperature $T_c$.

On the other hand, in the past years many activities have been
devoted to the domain-wall dynamics
\cite{lyu99,nat01,gla03,mig04,bra05,kle06,kle07}. For magnetic
materials, for example, a domain wall separates domains with
different spin orientations. Microscopically, the domain wall may
move and create bubbles, and macroscopically, it may propagate and
roughen. At the zero temperature, there occurs a pinning-depinning
phase transition induced by quenched randomness
\cite{now98,lyu99,rot01}. For a magnetic system with weak
disorder, the domain wall does not propagate unless the external
magnetic field $h$ exceeds a threshold $h_c$. At the critical
field $h_c$, a roughening phenomenon is also observed
\cite{mjos96}. When a periodic external field $h(t)=h_0\cos(\omega
t)$ is applied, the second-order phase transition is softened to a
hysteresis loop \cite{lyu99,nat01,gla03,bra05}. Most these works
concerning the domain-wall motion concentrate on the stationary
state at the zero or very low temperatures and in response to the
external magnetic field $h(t)$.

In this paper, we systematically investigate the dynamic
relaxation with a domain wall at the critical temperature. For
simplicity, we assume that no external field is applied.
Macroscopically, therefore, the domain wall does not propagate. We
should only keep in mind that different from the case at the zero
temperature, here the bulk also evolves in time. To be specific,
we consider the dynamic relaxation starting from a {\it
semi-ordered} initial state. For the Ising model, for example, the
semi-ordered initial state consists of two fully-ordered domains
with opposite spin orientations. As time evolves, the domain wall
roughens, and looks like a growing interface. In this paper, we
call it the domain interface. Such a domain-wall dynamics is
theoretically and practically important. Theoretically, it is very
interesting to investigate the short-time dynamic scaling behavior
starting from the semi-ordered state, in comparison with that
starting from the ordered or random state. It extends the study of
the domain-wall motion at the zero or very low temperatures to the
critical temperature, and especially explores the dynamic behavior
far from equilibrium. In this paper, we intend to clarify first
the dynamic scaling behavior of model A \cite{hoh77}. Then the
dynamic theory may be generalized to model B. Along this
direction, one might find the way to study relevant dynamic
processes of driven diffusive lattice gases \cite{alb02,car04}.

Furthermore, the non-equilibrium critical dynamics around a
surface is also an important topic
\cite{rit95,ple04,ple04b,ple05}. For the dynamic relaxation
starting from the random state, the dynamic evolution of the
magnetization at surface is controlled by both the scaling
dimension $x_0$ of the global initial magnetization and the static
exponent $\beta_1$ of the surface magnetization
\cite{rit95,ple04}. For the dynamic relaxation starting from the
ordered state, it is expected that $\beta_1$ is sufficient to
describe the dynamic evolution of the magnetization at surface.
For the dynamic relaxation {\it without a surface} but starting
from the semi-ordered state, it looks somewhat like that there
exists a fictitious surface. The dynamic evolution of the
magnetization inside the domain interface is governed by an
exponent $\beta_1$. But this $\beta_1$ does not correspond to a
static exponent in equilibrium, and it is induced by the
semi-ordered initial state. Therefore, an additional purpose of
this paper is to compare the dynamic relaxation starting from the
semi-ordered state with that starting from the ordered state but
around a surface.

In Ref. \cite{zho07}, brief results on the dynamic relaxation of
the magnetization have been reported for the two-dimensional Ising
model. This paper aims at a comprehensive study of the topic, and
explores especially the dynamic scaling behavior of the Binder
cumulant (or susceptibility), height function and roughness
function. Furthermore, Monte Carlo simulations are performed also
for the three-dimensional Ising model, to study the dimension
dependence of the scaling functions and critical exponents as well
as the corrections to scaling. In Sec. II, the models and scaling
analysis are described, and in Sec. III, the numerical results are
presented. Finally, Sec. IV includes the conclusions.

\section{Model and scaling analysis}

\subsection{Model}

The $d$-dimensional Ising model is the simplest model for magnetic
materials, exhibiting a second-order phase transition. The
Hamiltonian is written as
\begin{equation}
-\frac{1}{kT}H= K  \sum_{<ij>}  S_i S_j\ , \label{equ5}
\end{equation}
where $S_i =\pm 1$ is an Ising spin at site $i$ of a square or
cubic lattice, the sum is over the nearest neighbors, and $T$ is
the temperature. In this paper, we set the temperature at its
critical value $T_c$; or in other words, we set $K$ at its
critical value $K_c$. The Hamiltonian of the Ising model itself
does not include an intrinsic dynamics. For example, Monte Carlo
algorithms may be introduced to simulate the dynamic evolution of
the system. It is generally believed that the Monte Carlo dynamics
is in the same universality class of the Langevin equation.

Let us consider a kind of dynamic relaxation processes at the
critical temperature. After a macroscopic initial state at very
low temperatures is prepared, the dynamic system is suddenly
quenched to the {\it critical} temperature, and then released to
the dynamic evolution of model A \cite{hoh77,zhe98}. For the
dynamics of model A, the order parameter and other relevant
physical quantities are not conserved during the dynamic
evolution. In Monte Carlo simulations, it can be simply realized
with a standard one-spin flip. In this paper, the heat-bath
algorithm is always used in the dynamic Monte Carlo simulations.
Selecting a single spin $S_i$, we flip it with the transition
rate,
\begin{equation}
   P(S_i \to S^{'}_i) \sim
   \frac{exp(KS^{'}_i\sum_{j(i)}S_j)}{exp(K\sum_{j(i)}S_j)+exp(-K\sum_{j(i)}S_j)},
\label{equ8}
\end{equation}
where $j(i)$ labels the nearest neighbors of the site $i$, and $c$
is the normalization constant. In fact, other Monte Carlo
algorithms, such as Metropolis algorithms, Monte Carlo algorithms
with a multi-spin flip and rejection-free Monte Carlo algorithms
etc, yield the same results. The condition is that the algorithms
should be {\it local}, i.e., only spins in a local region are
flipped in a single flip.

With Monte Carlo simulations, we first study the critical
relaxation starting from a {\it semi-ordered} state, taking the
two-dimensional ($2$D) and three-dimensional ($3$D) Ising model as
examples. The Ising  model is defined on a rectangular lattice $2L
\times L$ in two dimensions and $2L \times L^2$ in three
dimensions, with a linear size $2$L in the $x$ direction and $L$
in the other directions. Periodic boundary conditions are used in
all the directions. The semi-ordered state is such a state, that
spins are positive on the sublattice $L^d$ ($d=2$ or $3$) at the
right side and negative on the sublattice $L^d$ at the left side.
For convenience, we set the $x$-axis such that the domain wall
between the positive and negative spins is located at $x = 0$. So
the $x$ coordinate of a lattice site is a half-integer.

\begin{figure}[p]
\epsfysize=6.cm \epsfclipoff \fboxsep=0pt
\setlength{\unitlength}{1.cm}
\begin{picture}(6,6)(0,0)
\put(-1.0,3.3){{\epsffile{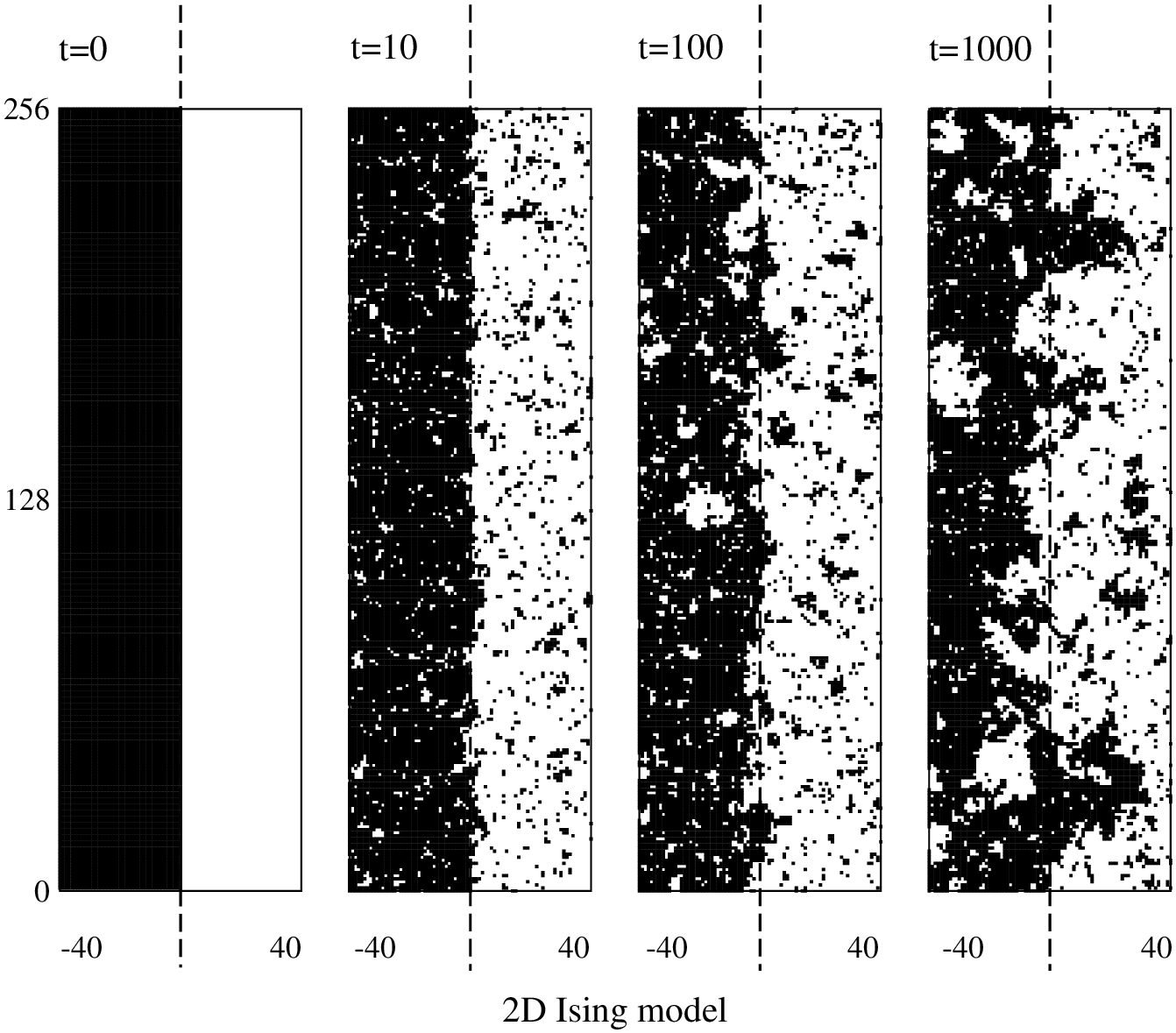}}}
\end{picture}
\caption{Dynamic relaxation from a semi-ordered state is simulated
for the $2$D Ising model at the critical temperature. The spin
configuration of the domain interface is shown in a spatial widow
$[-40 , 40]$ at the time $t = 0, 10, 100, 1000$ (from left to
right). Black points denote $S_i = -1$ and white points denote
$S_i = 1$. The lattice size $L = 256$ is used in the Monte Carlo
simulations.} \label{f10}
\end{figure}

After preparing the semi-ordered initial state, we update the
spins with the heat-bath algorithm at the critical temperature
$T_c$. Since no external magnetic field is added, macroscopically
the domain wall does not move. As time evolves, however, the
domain wall fluctuates and creates bubbles. As a result, the
domain wall becomes thicker and thicker, and a dynamic {\it
roughening} process occurs. Therefore, we call it {\it a domain
interface}. In Fig.~\ref{f10}, the dynamic evolution of the spin
configuration around the domain wall is illustrated. Somewhat
different from a standard growing interface, here the bulk evolves
in time. In analyzing the dynamic properties of the domain
interface, this must be kept in mind.

For comparison, we also perform Monte Carlo simulations of the
Ising model with a {\it free} or {\it disordered} surface, but
starting from the {\it ordered} state. In this case, the lattice
is taken to be $L^d$ ($d=2$ or $3$). For the free surface, a free
boundary condition is used in the $x$ direction, while periodic
boundary conditions are used in other directions. For convenience,
we set the $x$-axis such that the free surface locates at $ x = 1$
or $ L $. For the disordered surface, the spin $S_i$ on the
surface couples to a random field $h_i$ through the interaction
$-H_i/kT = Kh_iS_i$ with the random filed $h_i = \pm 1$. The
disordered boundary condition is only implemented in the $x$
direction, and periodic conditions are used in other directions.
Since the initial state is the ordered state, the magnetization
decays in time. In particular, the magnetization is also
$x$-dependent due to the geometric surface. The dynamic behavior
of the magnetization around the surface is governed by the surface
exponents, while that at bulk is controlled by the bulk exponents.
The region affected by the geometric surface grows in time, and it
looks like that the surface becomes thicker and thicker.
Phenomenologically, this dynamic behavior is similar to that of
the domain interface.

Finally, to expose the dynamic evolution of the {\it bulk}, we
perform Monte Carlo simulations of the Ising model with periodic
boundary conditions in {\it all} directions, starting from the
{\it ordered} state. The lattice is taken to be $L^d$ in $d$
dimensions.

\subsection{Scaling analysis}

We first analyze the dynamic scaling behavior of the domain
interface. Due to the semi-ordered initial state, the time
evolution of the dynamic system is inhomogeneous in the $x$
direction. Therefore, we measure the magnetization and its second
moment as functions of $x$ and $t$. In two dimensions, for
example,
\begin{equation}
M^{(k)}(t,x) = \frac{1}{L^k} \left \langle \left [\sum^L_{y =
1}S_{xy}(t)\right ]^k \right \rangle,  \quad k = 1,2.
\label{equ10}
\end{equation}
Here $S_{xy}(t)$ is the spin at the time $t$ on the lattice site
$(x,y)$, $L$ is the lattice size, and $<\ldots>$ represents the
statistical average. For convenience, we also use $M(t,x) \equiv
M^{(1)}(t,x)$ to denote the magnetization. Then we can define a
time-dependent Binder cumulant \cite{zhe98,zho07},
\begin{equation}
U(t,x) = \frac {M^{(2)}(t,x)}{M(t,x)^{2}} - 1. \label{equ30}
\end{equation}
The susceptibility $M^{(2)}(t,x)-M(t,x)^{2}$ or the Binder
cumulant $U(t,x)$ describes the fluctuation in the $y$ direction.
In three dimensions, we simply use $S_{xyz}$ to denote the spin on
the lattice site $(x,y,z)$, and similarly define the magnetization
and Binder cumulant.

In order to characterize the growth of the domain interface and
its fluctuation in the $x$ direction, we introduce a height
function and its second moment in the $x$ direction,
\begin{equation}
h^{(k)}(t) = \frac{1}{L^k} \left \langle \left [\sum^{L}_{x =
1}S_{xy}(t)\right ]^k \right \rangle, \quad k = 1,2. \label{equ40}
\end{equation}
Here $<\ldots>$ represents not only the statistic average but also
the average in the $y$ direction. As usual, we also use the
notation $h(t) \equiv h^{(1)}(t)$. Then the roughness function of
the domain interface is defined as
\begin{equation}
\omega^2(t) = h^{(2)}(t) - h(t)h(t). \label{equ50}
\end{equation}
Except for the scaling dimension of the magnetization, the height
function measures the thickness of the domain interface, while the
roughness function represents the fluctuation of the domain
interface.

At the critical temperature $T_c$ and in the thermodynamic limit,
there are two length scales in the dynamic system, i.e., $x$ and
the non-equilibrium spatial correlation length $\xi(t)$. For a
finite system, the lattice size $L$ is an additional length scale.
In general, one may believe that $\xi(t)$ universally grows as
$\xi (t) \sim t^{1/z}$ in all spatial directions, because of the
homogeneity of the interactions in the Hamiltonian. Therefore,
general scaling arguments lead to the scaling form of the
magnetization and its second moment
\begin{equation}
M^{(k)}(t,x,L)=t^{-k\beta /\nu z}
\widetilde{M}^{(k)}(t^{1/z}/x,t^{1/z}/L), \quad k = 1,2.
\label{equ55}
\end{equation}
Here $\beta$ and $\nu$ are the static exponents, and $z$ is the
dynamic exponent. On the right side of the equation, the overall
factors $t^{-k\beta / \nu z}$ indicates the scaling dimension of
$M^{(k)}$, and the scaling function
$\widetilde{M}^{(k)}(t^{1/z}/x,t^{1/z}/L)$ represents the scale
invariance of the dynamic system. In general, the scaling form in
Eq.~(\ref{equ55}) holds already in the {\it macroscopic}
short-time regime, after a microscopic time scale $t_{mic}$
\cite{jan89,zhe98}.

For the magnetization, the scaling function
$\widetilde{M}(t^{1/z}/x,t^{1/z}/L)$ is independent of $L$ in the
thermodynamic limit $L \to \infty$. Then the scaling form is
simplified to
\begin{equation}
M(t,x)=t^{-\beta /\nu z}\widetilde{M}(t^{1/z}/x). \label{equ60}
\end{equation}
For the susceptibility, it is different. For a sufficiently large
lattice and in the short-time regime, the non-equilibrium spatial
correlation length $\xi(t)$ is much smaller than the lattice size
$L$. Therefore, the spatially correlating terms $<S_{xy_1}
S_{xy_2}>-<S_{xy_1}><S_{xy_2}>$ with $|y_2-y_1|>\xi(t)$ can be
neglected. In other words, one of the two summations over $y_1$
and $y_2$ in the susceptibility $M^{(2)}(t,x)-M(t,x)^{2}$ is
suppressed. It then leads to the finite-size behavior
$M^{(2)}(t,x)-M(t,x)^{2} \sim 1/L^{d-1}$ ($d=2$ or $3$)
\cite{zhe98}. Together with Eqs.~(\ref{equ55}) and (\ref{equ60}),
one may derive the scaling form of the Binder cumulant
\cite{zhe98}
\begin{equation}
U(t,x) = \frac{t^{(d - 1)/z}}{L^{d-1}}\widetilde{U}(t^{1/z}/x).
\label{equ70}
\end{equation}
The Binder cumulant is interesting, for only the dynamic exponent
$z$ is involved.

By definition, the height function $h(t)$ is nothing but the
average magnetization in the positive domain, i.e.,
$h(t)=\sum_{x>0} M(t,x)/L$. In general, $h(t)$ does not obey a
simple power law. Its behavior replies on the scaling function
$\widetilde{M}(t^{1/z}/x)$. In fact, one may deduce a scaling form
for $h(t)$ from Eq.~(\ref{equ55}),
\begin{equation}
h(t) = t^{-\beta/\nu z}\widetilde{h}(t^{1/z}/L). \label{equ75}
\end{equation}
Different from $M(t,x)$, here one should not ignore the dependence
on the lattice size $L$, for the scaling function
$\widetilde{h}(t^{1/z}/L)$ just represents the dynamic effect of
the domain interface. This is obvious from the definition
$h(t)=\sum_{x>0} M(t,x)/L$. On the other hand, one should also
note that the height function here is scaled as the magnetization,
not a spatial length.

Similar to Eq.~(\ref{equ75}), one may also assume that the scaling
form for the roughness function is
$\omega^2(t)=t^{-2\beta/\nu/z}F(t^{1/z}/L)$. For later
convenience, we separate a factor $t^{1/z}/L$ from $F(t^{1/z}/L)$,
and rewrite the scaling form as
\begin{equation}
\omega^2(t) = \frac{t^{(1 -2\beta/\nu)/z}}{L}
\widetilde{\omega^2}(t^{1/z}/ L ). \label{equ80}
\end{equation}
In general, $\omega^2(t)$ does not exhibit a power-law behavior.
This is different from a standard growing interface. The reason is
that here $\omega^2(t)$ includes fluctuations from the domain
interface and the bulk. In fact, we will show in the next section
that the scaling function $\widetilde{\omega^2}(t^{1/z}/ L)$
describes the fluctuation induced by the domain interface.

The scaling forms in Eqs.~(\ref{equ55})$-$(\ref{equ80}) can be
also applied to the dynamic relaxation with a free or disordered
surface, but starting from the {\it ordered} state. One should
only keep in mind that $\beta$, $\nu$ and $z$ are the critical
exponents {\it at bulk}. The critical exponents at surface should
be deduced from the scaling functions in
Eqs.~(\ref{equ55})$-$(\ref{equ80}). In this case, the lattice is
taken to be $L^d$. The inhomogeneity in the $x$ direction is
induced by the surface.

The purpose of this paper is to investigate whether the scaling
forms in Eqs.~(\ref{equ55})$-$(\ref{equ80}) do hold in the dynamic
relaxation with the domain interface and with the free or
disordered surface. With Monte Carlo simulations, we study
characteristics of the scaling functions, and extract
corresponding critical exponents. Dynamic systems with the domain
interface and with the free or disordered surface share some
common features, although they are intrinsically different. The
domain interface is induced by the geometric structure of the
semi-ordered initial state, while the dynamic relaxation with the
free or disordered surface is controlled by the geometric surface
which remains even in equilibrium.

It is important that the height function $h(t)$ and roughness
function $\omega^2(t)$ in Eqs.~(\ref{equ40}) and (\ref{equ50})
include the dynamic evolution of the bulk. Therefore, their
behaviors deviate from those of a standard growing interface. To
obtain the dynamic features of a pure growing interface such as
the power-law behavior and the roughness exponent etc, we need to
subtract the contribution of the bulk. Therefore, we finally
perform Monte Carlo simulations of the Ising model on a lattice
$L^d$ with periodic boundary conditions in all directions, and
starting from the ordered state. In this case, the dynamic system
is homogeneous in all directions. The height function $h_b(t)$ and
the roughness function $\omega_b^2(t)$ are just the line
magnetization and its susceptibility in the $x$ or $y$ direction.
The scaling functions $\widetilde{h}(t^{1/z}/L)$ and
$\widetilde{\omega^2}(t^{1/z}/ L )$ in Eqs.~(\ref{equ75}) and
(\ref{equ80}) are constants. In other words, $h_b(t)$ and
$\omega_b^2(t)$ show a power-law behavior.

Then we may redefine the pure height function and roughness
function for the dynamic relaxation of the domain interface or
around the surface by subtracting the contribution from the bulk
\begin{equation}
Dh(t,L) = h_b(t)-h(t), \label{equ85}
\end{equation}
\begin{equation}
D\omega^2(t,L) = \omega^2(t)-\omega_b^2(t). \label{equ87}
\end{equation}
We may expect that $Dh(t,L)$ and $D\omega^2(t,L)$ exhibit a
power-law behavior as in the case of a standard growing interface.
Here we should note that we define $Dh(t,L)$ as $h_b(t)-h(t)$
rather than $h(t)-h_b(t)$, for $h(t)$ decays in time faster than
$h_b(t)$.

\section{Monte Carlo Simulation}

For the $2$D Ising model, our main results are presented with $L =
512$ at $K_c = 0.44069$, and the maximum updating time is $t_M =
25600$. Additional simulations with $L = 1024$ and $L = 256$ are
performed, to investigate the finite-size scaling behavior and
finite-size effect. The total samples for average are $24000$. For
the $3$D Ising model, the main results are obtained with $L = 128$
at $K_c = 0.22165$, and the maximum updating time is $t_M = 2560$.
Additional simulations with $L = 256$ and $L = 64$ are performed
to investigate the finite-size scaling behavior and finite-size
effect. The total samples for average are $30000$. The statistical
errors are estimated by dividing the total samples into two or
three subgroups. If the fluctuation in the time direction is
comparable with or larger than the statistical error, it will be
taken into account.

Theoretically, the scaling forms described in the previous section
hold in the macroscopic short-time regime, after a microscopic
time scale $t_{mic}$. $t_{mic}$ is not universal, and relies on
microscopic details of the dynamic systems. In Monte Carlo
simulations, for example, $t_{mic}$ is typically tens or hundreds
of Monte Carlo time steps \cite{zhe98}. With quenched disorder or
frustration, $t_{mic}$ could be longer. For the simple Ising model
with the nearest neighbor interactions, $t_{mic}$ is rather short,
about $10$ - $20$ time steps. Therefore, critical exponents are
typically obtained in the time intervals $[20, 25600]$ in two
dimensions and $[10, 2560]$ in three dimensions. From the data
collapse of different $x$ and $L$, one may observe the scaling
functions in a even larger time window.

\subsection{Magnetization}

\begin{figure}[p]
\epsfysize=6.cm \epsfclipoff \fboxsep=0pt
\setlength{\unitlength}{1.cm}
\begin{picture}(6,6)(0,0)
\put(-1.0,0.3){{\epsffile{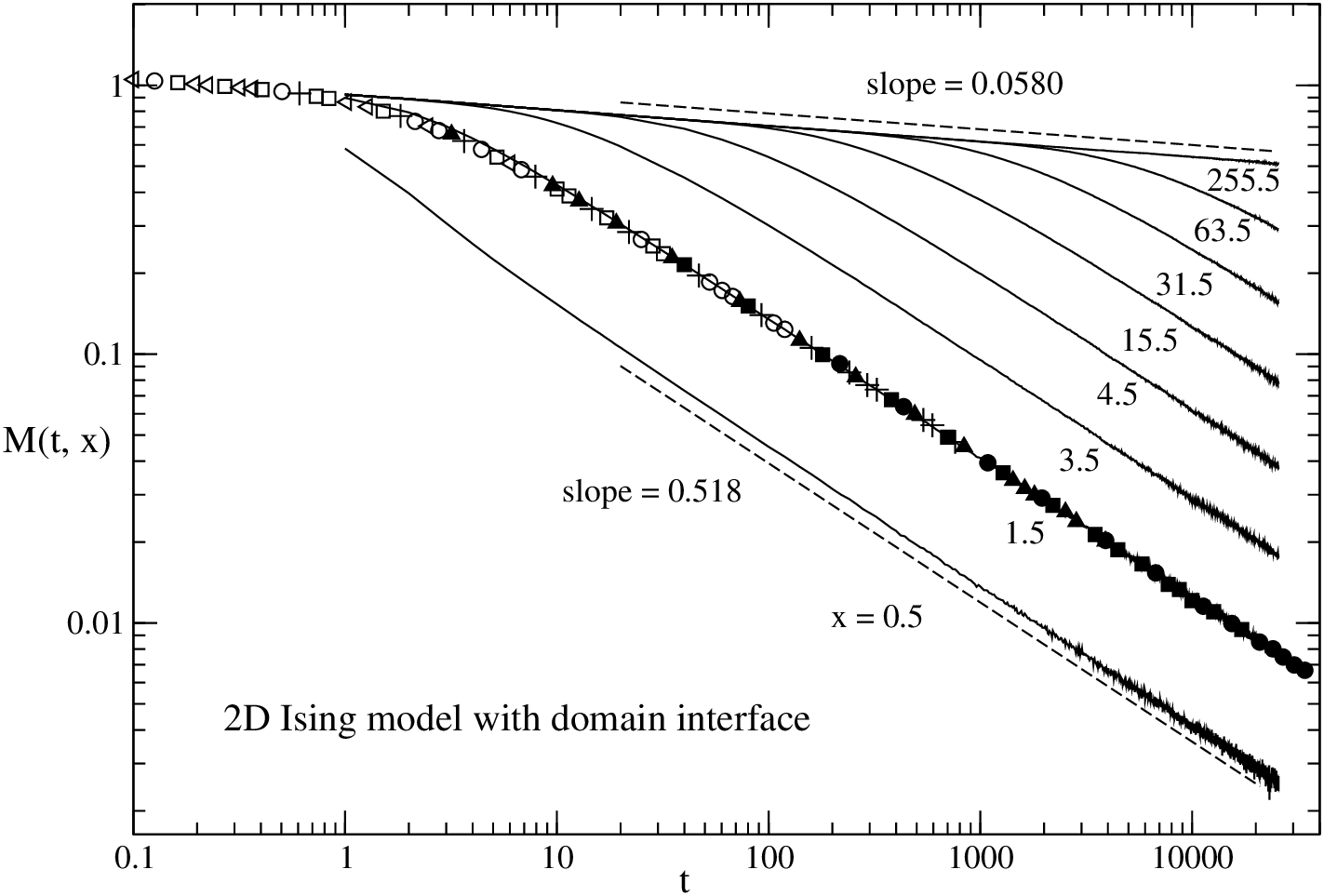}}}
\end{picture}
\caption{The magnetization of the $2$D Ising model starting from
the semi-ordered state is plotted with solid lines on a double-log
scale. Dashed lines show the power-law fits. According to
Eq.~(\ref{equ60}), data collapse for different $x$ is demonstrated
at a fixed $x' = 1.5$. Solid circles, solid squares, solid
triangles, pluses, open circles, open squares and open triangles
correspond to $x = 0.5, 1.5, 3.5, 7.5, 15.5, 31.5$ and $63.5$
respectively. From Ref.~\cite{zho07}, this figure is reproduced by
permission of Europhys. Lett..} \label{f0}
\end{figure}

The time evolution of the magnetization of the $2$D Ising model
starting from the semi-ordered state is displayed in
Fig.~\ref{f0}. For a sufficiently small $s$, e.g., $x=255.5$ and
$t<t_M=25600$, $M(t,x)$ approaches the non-linear decay {\it at
bulk}, $M(t,x) \sim t^{-\beta/\nu z}$ \cite{zhe98}. The exponent
$\beta/\nu z = 0.0580(3)$ measured from the slope of the curve is
well consistent with $\beta=1/8$, $\nu=1$ and $z=2.16(1)$ reported
in the literature \cite{zhe98}. For a sufficiently large $s$,
e.g., $x=0.5$ and $t>20$, $M(t,x)$ exhibits also a power-law
behavior, but decays {\it much faster} than at bulk. In other
words, we catch some features of the scaling function
$\widetilde{M}(s)$ in Eq.~(\ref{equ60}),
\begin{equation}
   \widetilde{M}(s) \sim \{
   \begin{array}{lll}
    \mbox{const}    & \quad &  \mbox{$s \to  0$ } \\
   s^{-\beta_0 / \nu}  & \quad &  \mbox{$s \to  \infty$}
   \end{array},
\label{equ110}
\end{equation}
with $s=t^{1/z}/x$. In the limit $s \to \infty$, one may define an
{\it interface} exponent $\beta_1$ such that
\begin{equation}
M(t,x) \sim t^{-\beta_1/\nu z}\cdot x^{\beta_0 / \nu}, \quad
\beta_1=\beta+\beta_0. \label{equ115}
\end{equation}
Inside the interface, the power-law decay of the magnetization is
governed by the interface exponent $\beta_1$, while outside the
interface, it is controlled by the bulk exponent $\beta$. In
Fig.~\ref{f0}, one measures $\beta_1/\nu z=0.518(4)$, and then
calculates $\beta_1=1.119(9)$ and $\beta_0/\nu=0.994(9)$. Similar
to the exponent $x_0$ in the dynamic relaxation starting from the
random state \cite{jan89,zhe98}, $\beta_0$ here is induced by the
semi-ordered initial state. Accounting the error, one may believe
$\beta_0/\nu =1$, which suggests that $M(t,x)$ is an analytic
function of $x$. This result is also supported by the simulations
of the $3$D Ising model. Since $\beta_1$ is much bigger than
$\beta$, the magnetization inside the domain interface decays much
faster than that at bulk. This phenomenon is understandable, for
the dynamic evolution of the spins in the positive domain is
strongly affected by those in the negative domain, and vice versa.

\begin{figure}[p]
\epsfysize=6.cm \epsfclipoff \fboxsep=0pt
\setlength{\unitlength}{1.cm}
\begin{picture}(6,6)(0,0)
\put(-1.0,0.3){{\epsffile{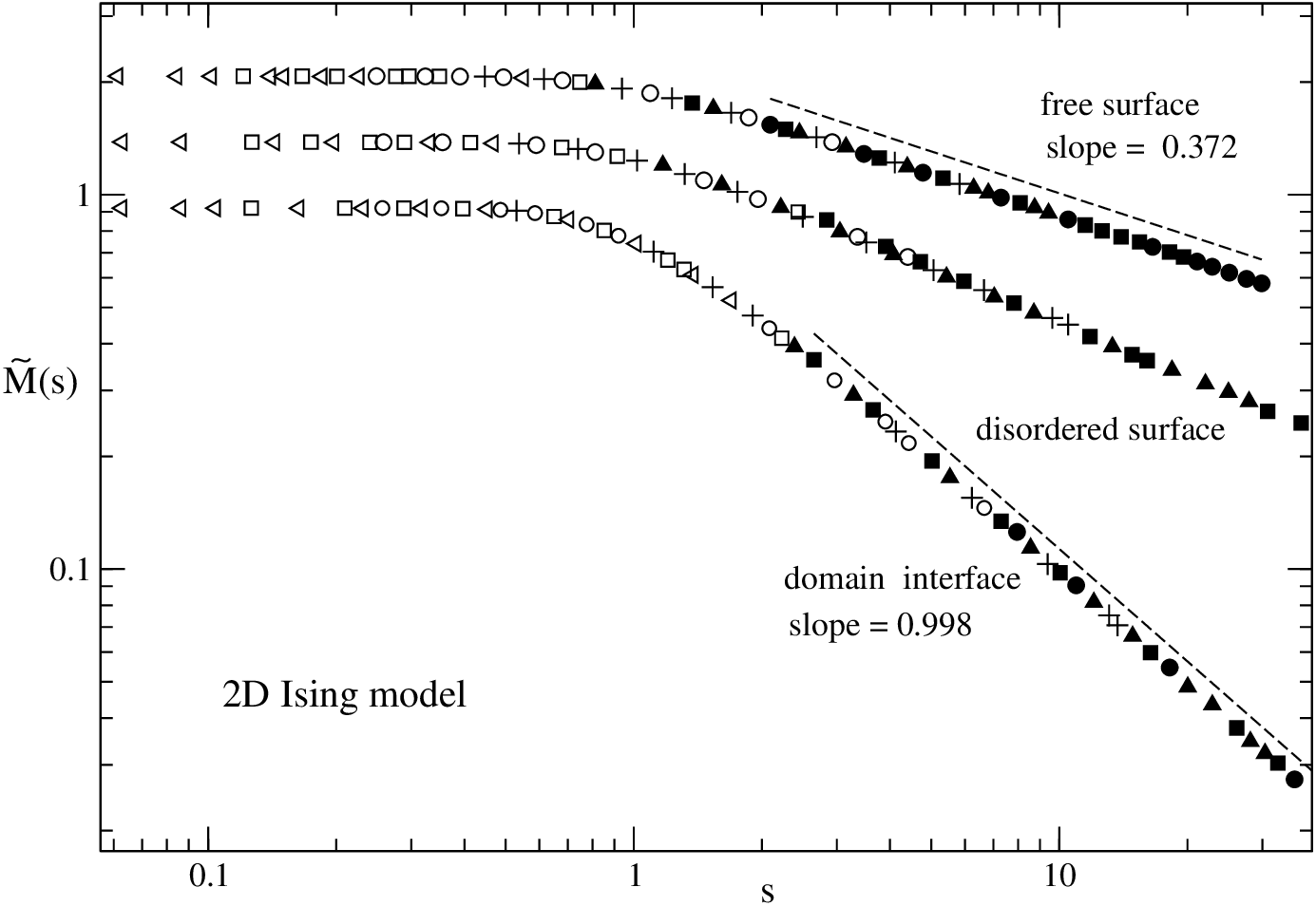}}}
\end{picture}
\caption{The scaling functions $\widetilde{M}(s)$ with
$s=t^{1/z}/x$ is plotted on a double-log scale, for the $2$D Ising
model with the free surface, disordered surface and domain
interface (from above). Data collapse for different $x$ is
observed. Dashed lines show the power-law fits. From
Ref.~\cite{zho07}, this figure is reproduced by permission of
Europhys. Lett..} \label{f3}
\end{figure}

To fully confirm the scaling form in Eq.~(\ref{equ60}), for
example, we fix $x' =1.5$, and change the time scale $t$ of
another $x$ to $(x'/x)^z\ t$, and the scale of $M(t,x)$ to
$(x'/x)^{-\beta/\nu}M(t,x)$. As shown in Fig.~\ref{f0}, all data
of different $x$ nicely collapse to the curve of $x'=1.5$. This
validates Eq.~(\ref{equ60}). Alternatively, we may plot $t^{\beta/
\nu z} M(t,x)$ as a function of $s=t^{1/z}/x$. According to
Eq.~(\ref{equ60}), all data of different $x$ should collapse onto
the master curve $\widetilde{M}(s)$. This is shown in
Fig.~\ref{f3}. Clearly, $\widetilde{M}(s) \to const$ when $s \to
0$, while $\widetilde{M}(s) \to s^{-\beta_0/\nu}$ when $\ s \to
\infty$.

\begin{figure}[p]
\epsfysize=6.cm \epsfclipoff \fboxsep=0pt
\setlength{\unitlength}{1.cm}
\begin{picture}(6,6)(0,0)
\put(-1.0,0.3){{\epsffile{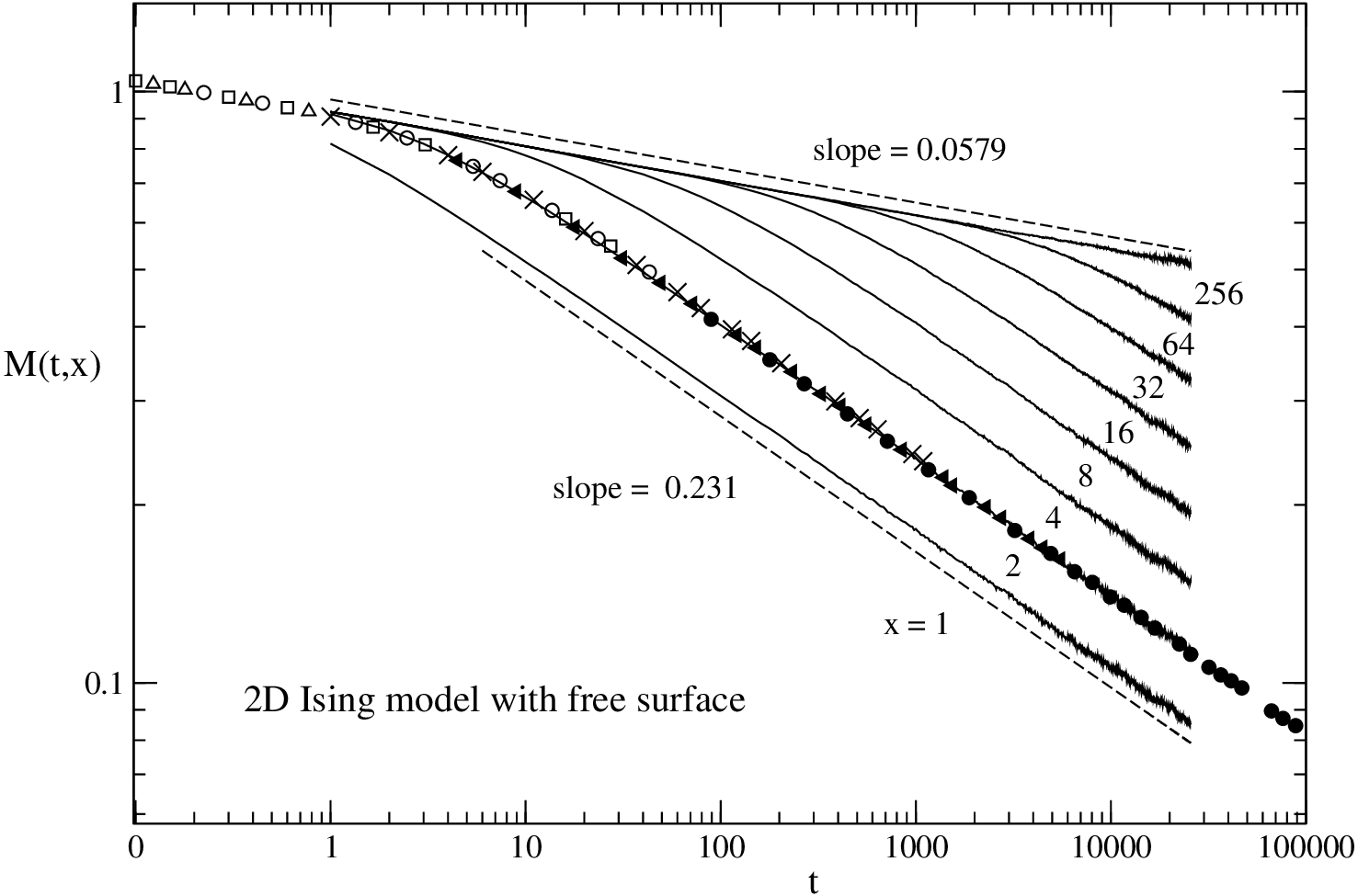}}}
\end{picture}
\caption{The magnetization of the $2$D Ising with a free surface
and starting from an ordered state, is plotted with solid lines on
a double-log scale. Dashed lines show the power-law fits.
According to Eq.~(\ref{equ60}), data collapse for different $x$ is
demonstrated at a fixed $x' = 2$. Solid circles, solid triangles,
pluses, open circles, open squares and open triangles correspond
to $x = 1, 4, 8, 16, 32 $ and $64$ respectively.} \label{f1}
\end{figure}

For comparison, the time evolution of the magnetization of the
$2$D Ising model with a free surface but starting from the ordered
state is shown in Fig.~\ref{f1}. For a sufficiently small $s$,
e.g., $x = 256$ and $t<t_M=25600$, $M(t,x)$ approaches also the
non-linear decay {\it at bulk}, $M(t,x) \sim t^{-\beta/\nu z}$
with $\beta/\nu z = 0.0579(4)$. For a sufficiently large $s$,
e.g., $x= 1$ and $t>20$, $M(t,x)$ exhibits also a power-law
behavior. Assuming again the scaling ansatzes in
Eqs.~(\ref{equ110}) and (\ref{equ115}), the measurement of the
slope yields $\beta_1 / \nu z = 0.231(1)$. Then one calculates
$\beta_1 = 0.499(2)$. It is in good agreement with the surface
exponent $\beta_s = 1/2 $ for the free surface \cite{ple04a}. Now,
the exponent $\beta_0/\nu$ is estimated to be $0.374$. Therefore,
$M(t,x)$ is {\it not} an analytic function of $x$, when $x$
approaches the free surface. This is very different from the
domain interface. Finally, the data collapse according to
Eq.~(\ref{equ60}) is also shown in Fig.~\ref{f1}, and it fully
confirms the scaling form.

As in Figs.~\ref{f0} and \ref{f1}, similar analysis can be carried
out for the magnetization of the $2$D Ising model with a
disordered surface. For a small $s$, one measures $\beta/\nu z
=0.0578(7)$. For a large $s$, careful analysis shows that the
power-law behavior is not perfect \cite{zho07}. In the equilibrium
state, one may show that the surface exponent $\beta_s$ of the
disordered surface remains $1/2$, but with a logarithmic
correction to scaling \cite{ple04a}. Therefore, we fit the
time-dependent magnetization at $x = 0.5$ with a logarithmic
correction to scaling, i.e., $M(t)=c_1t^{- \alpha_1}/(1+ c_2
\ln(t))^{1/2}$, and derive $\alpha_1 = 0.231$, consistent with
$\beta_1 / \nu z = 0.231(1)$ for the free surface. If one fixes
$c_2=0$, it yields $\alpha_1=0.272$, significantly different from
$0.231(1)$.

In Fig.~\ref{f3}, the scaling function $\widetilde{M}(s)$ with
$s=t^{1/z}/x$ is plotted for the domain interface, free surface
and disordered surface. We clearly observe the characteristic of
the scaling function in Eq.~(\ref{equ110}), and measure the
exponent $\beta_0 = 0.998(5)$ for the domain interface, and
$0.372(6)$ for the free surface. Due to the logarithmic correction
to scaling, $\widetilde M(s)$ of the disordered surface decays
faster than that of the free surface at the large $s$ regime.

We emphasize that in the case of the free surface or disordered
surface, the exponent $\beta_1 \equiv \beta_s$ does describe the
critical behavior of the magnetization at the surface {\it in
equilibrium}. Around the free surface, for example, $M(\tau) \sim
(-\tau)^{\beta_1}$ with $\tau$ being the reduced temperature. It
is important that $\beta_1$ is induced by the geometric surface
which remains forever. In the case of the domain interface,
however, $\beta_1$ is induced by the geometric structure of the
semi-ordered initial state. When the dynamic system reaches its
equilibrium state, the influence of the initial state disappears
and the critical behavior of the magnetization is governed by the
bulk exponent $\beta$ {\it everywhere}. We should keep in mind,
however, that {\it exactly at the critical temperature}
(i.e.,$\tau=0$) and in the thermodynamic limit, the dynamic system
never reaches its equilibrium state in a finite time due to the
divergent correlating time. According to Eq.~(\ref{equ60}),
therefore, the domain interface and the free or disordered surface
behave similarly.

For the $3$D Ising model, the static and dynamic exponents at bulk
are known to be $\beta = 0.327(1), \nu = 0.630(2)$ and $z =
2.04(1)$ \cite{jas99}. For the free surface, the surface exponent
is $\beta_s = 0.795(10)$ \cite{bel06}. Following the procedure for
the $2$D Ising model, we have analyzed the scaling behavior of the
dynamic relaxation with the domain interface, free surface and
disordered surface. Especial attention is drawn to the critical
exponent $\beta_0/\nu$.

\begin{figure}[p]
\epsfysize=6.cm \epsfclipoff \fboxsep=0pt
\setlength{\unitlength}{1.cm}
\begin{picture}(6,6)(0,0)
\put(-1.0,0.3){{\epsffile{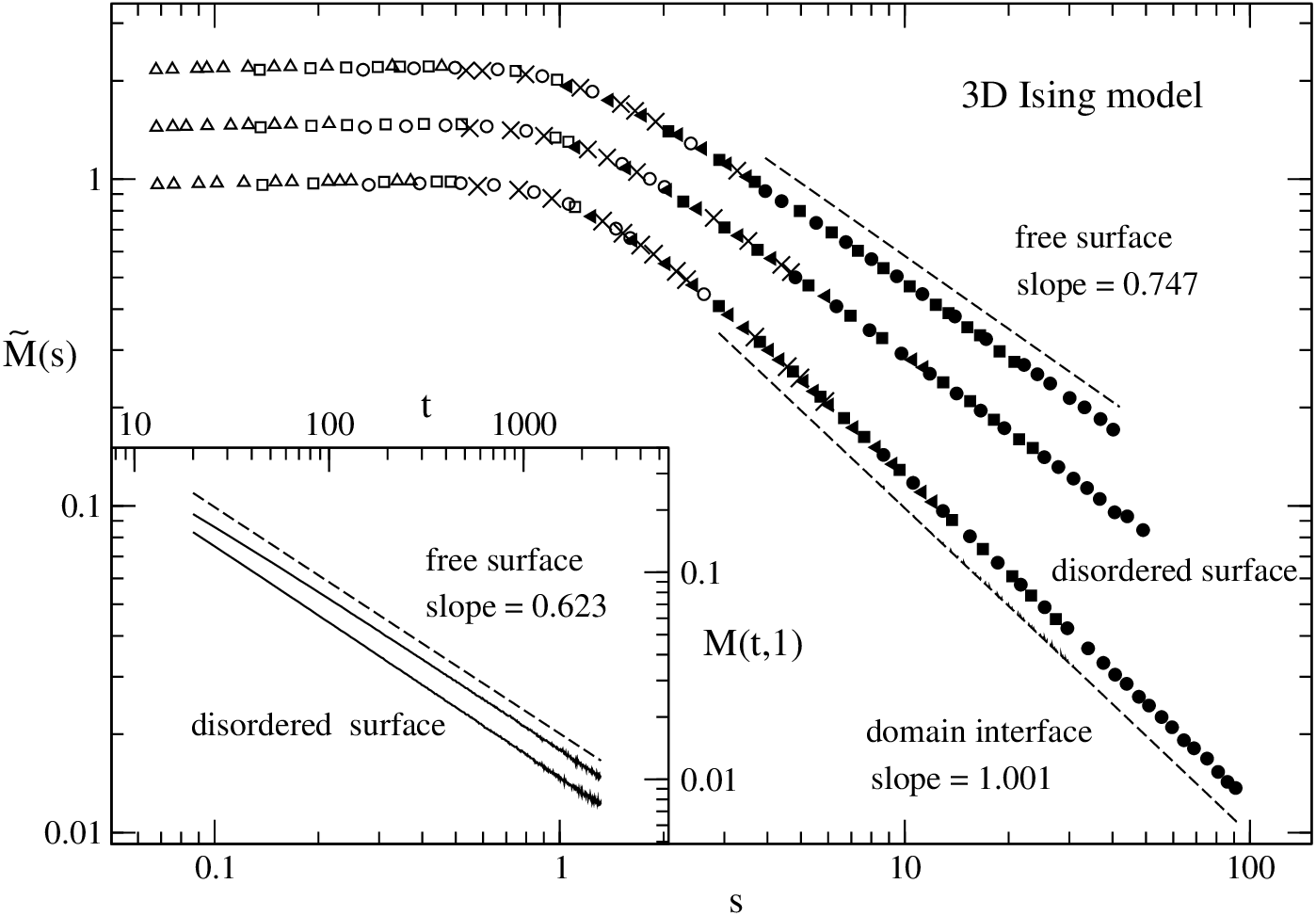}}}
\end{picture}
\caption{The scaling function $\widetilde{M}(s)$ with
$s=t^{1/z}/x$ is plotted on a double-log scale, for the $3$D Ising
model with the free surface, disordered surface and domain
interface (from above). Data collapse for different $x$ is
observed. Dashed lines show the power-law fits. No logarithmic
correction is detected for the disordered surface. In the inset,
the magnetization at $x = 1.0$ for the disordered surface and free
surface are shown.} \label{f5}
\end{figure}

Let us first consider the domain interface. For a small $s$, the
magnetization shows the power-law behavior at bulk, $M(t,x) \sim
t^{-\beta/\nu z}$. The critical exponent is estimated to be $\beta
/\nu z=0.253(5)$, well consistent with the value $0.253(1)$ at
bulk \cite{jas99}. For a large $s$, e.g., $x = 0.5$ and $t>10$,
the magnetization exhibits the power-law behavior $M(t,x) \sim
t^{-\beta_1/\nu z}$ with $\beta_1=\beta+\beta_0$ in
Eq.~(\ref{equ115}). From the slope of the curve, one obtains
$\beta_1/\nu z=0.744(2)$. Then one calculates the critical
exponent $\beta_0 / \nu = (0.744 - 0.253) \times 2.04 = 1.002(4)$.
For the $2$D Ising model, $\beta_0 / \nu = 0.998(5)$. These two
measurements of $\beta_0 / \nu$ strongly suggest $\beta_0 /
\nu=1$, and it is dimension-independent. Therefore $M(t,x)$ is an
analytic function of $x$. In Fig. \ref{f5}, the scaling functions
$\widetilde M(s)$ of the magnetization for the $3$D Ising model is
plotted. Data collapse for different $x$ is observed. From the
slope of the curve in the large $s$ regime, one measures $\beta_0
/ \nu=1.001(6)$.

Similar analysis is applied to the magnetization of the $3$D Ising
model with the free and disordered surfaces, and the scaling
function is also shown in Fig. \ref{f5}. Different from the case
of the $2$D Ising model, the large-$s$ tails of the scaling
function $\widetilde M(s)$ for the free and disordered surfaces
look parallel each other. In the inset of the figure, the
magnetization at $x = 1.0$ is displayed for both the free and
disordered surfaces. The slope is $\beta_1/\nu z=0.623(5)$ for the
free surface, and $0.632(2)$ for the disordered surface. The
difference is only one or two per cent, and the correction to
scaling is rather small. For the free surface, one estimates
$\beta_1=0.623(5)\times \nu z=0.801(6)$. Alternatively, one may
also measure $\beta_0 / \nu = 0.747(6)$ from the scaling function
obtained with different $x$, and then calculates $\beta_1 = \beta
+ \beta_0 = 0.798(4)$. These values of $\beta_1$ are well
consistent with the surface exponent $\beta_s=0.795(10)$
\cite{bel06}.

\subsection{Binder Cumulant}

\begin{figure}[p]
\epsfysize=6.cm \epsfclipoff \fboxsep=0pt
\setlength{\unitlength}{1.cm}
\begin{picture}(6,6)(0,0)
\put(-1.0,0.3){{\epsffile{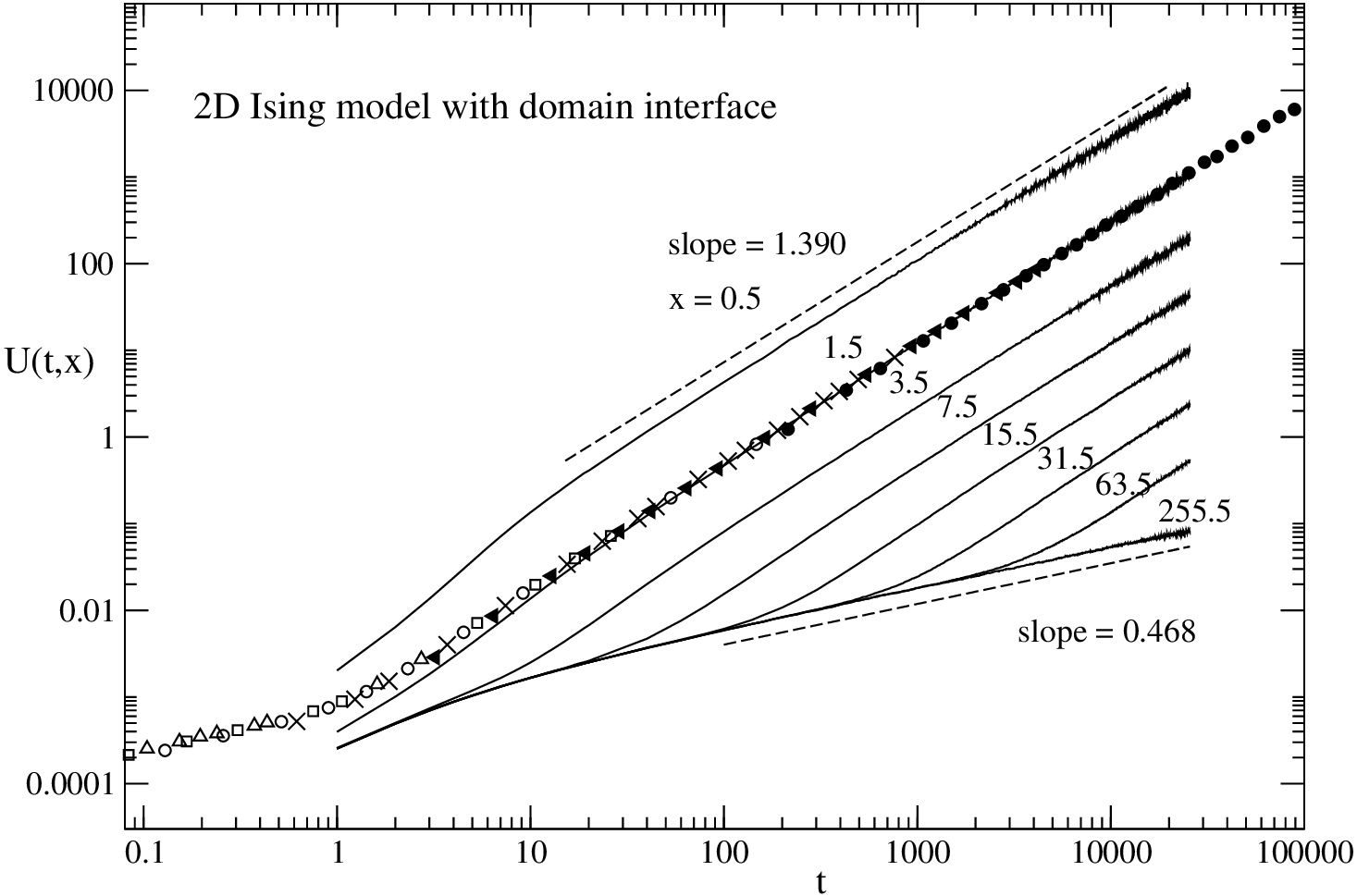}}}
\end{picture}
\caption{The Binder cumulant of the $2$D Ising model with the
domain interface is plotted with solid lines on a double-log
scale. Dashed lines show the power-law fits. According to
Eq.~(\ref{equ70}), data collapse for different $x$ is demonstrated
at a fixed $x' = 1.5$. Solid squares, solid triangles, pluses,
open circles, open squares and open triangles correspond to $x =
0.5, 3.5, 7.5, 15.5, 31.5$ and $63.5$ respectively. } \label{f6}
\end{figure}

In Fig. \ref{f6}, the time evolution of the Binder cumulant is
displayed for the $2$D Ising model starting from the semi-ordered
state. For a sufficiently small $s$, e.g., $x=255.5$ and
$t<t_M=25600$, the Binder cumulant exhibits the power-law behavior
at bulk, $U(t,x) \sim t^{(d - 1)/z}$. From the slope, one measures
$(d - 1)/z=0.468(4)$, and it is consistent with $(d -
1)/z=0.463(3)$ calculated from $z=2.16(1)$. For a sufficiently
large $s$, e.g., $x=0.5$ and $t>20$, the Binder cumulant grows
also by a power law, but much faster than that at the large $x$.
Then we extract the characteristic of the scaling function
$\widetilde{U}(s)$,
\begin{equation}
   \widetilde{U}(s) \sim  \{
   \begin{array}{lll}
    \mbox{const}    & \quad &  \mbox{$s \to  0$} \\
   s^{d_0}  & \quad &  \mbox{$s \to  \infty$}
   \end{array}.
   \label{equ160}
\end{equation}
In the limit $s \to \infty$, one may derive from
Eqs.~(\ref{equ70}) and (\ref{equ160}),
\begin{equation}
U(t,x) \sim t^{(d - 1 + d_0)/z}/(L^{d-1}x^{d_0}).
 \label{equ165}
\end{equation}
From the curve of $x = 0.5$ in Fig. \ref{f6}, one estimates $(d -
1 + d_0)/z=1.390(8)$. Taking $z = 2.16$ as input, one calculates
$d_0 = 1.390 \times 2.16 - (d - 1) = 2.00(2)$, very close to $2$.

\begin{figure}[p]
\epsfysize=6.cm \epsfclipoff \fboxsep=0pt
\setlength{\unitlength}{1.cm}
\begin{picture}(6,6)(0,0)
\put(-1.0,0.3){{\epsffile{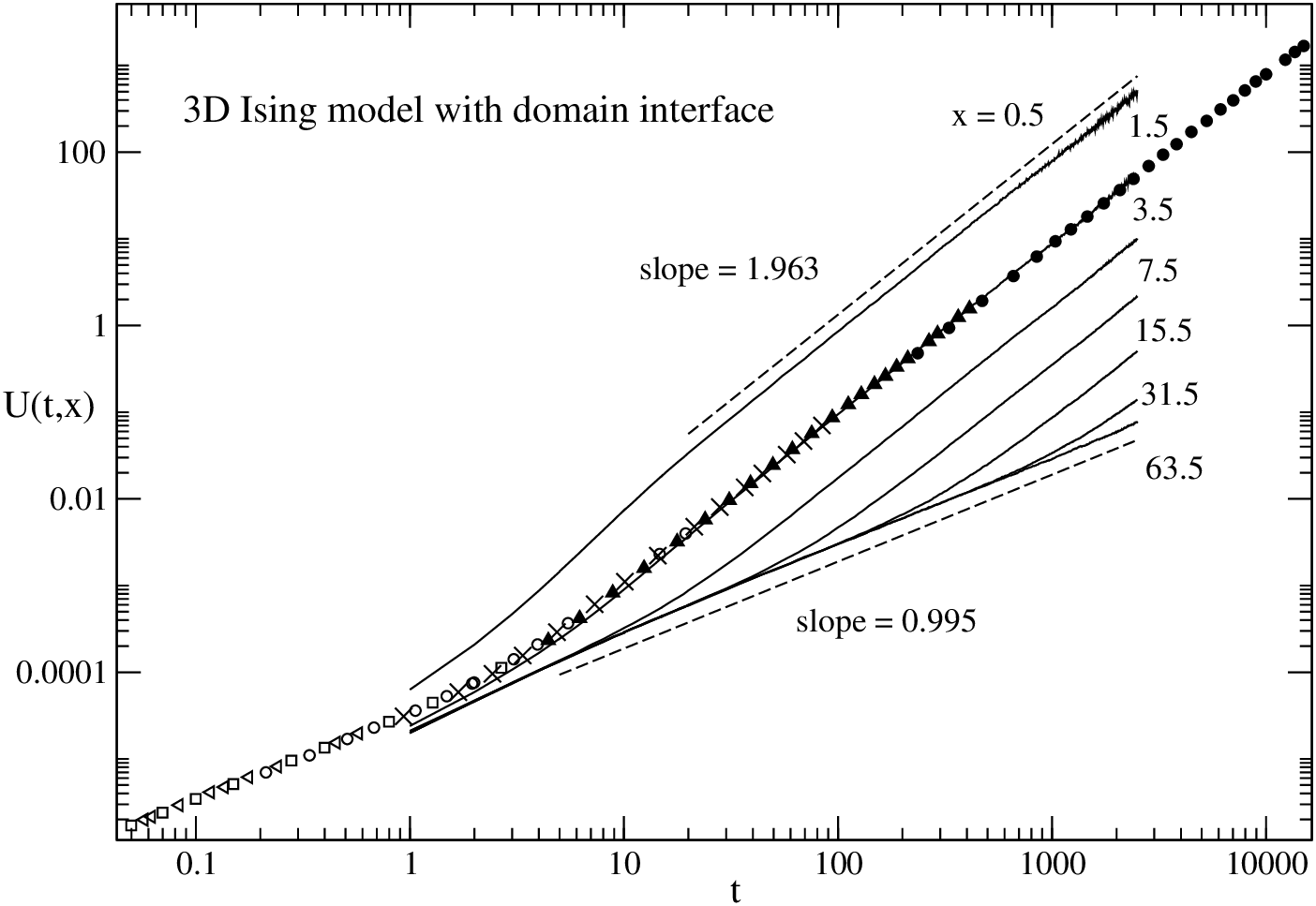}}}
\end{picture}
\caption{The Binder cumulant of the $3$D Ising model with the
domain interface is plotted with solid lines on a double-log scale
. Dashed lines show the power-law fits. According to
Eq.~(\ref{equ70}), data collapse for different $x$ is demonstrated
at a fixed $x' = 1.5$. Solid circles, solid triangles, pluses,
open circles, open squares and open triangles correspond to $x =
0.5 , 3.5, 7.5, 15.5, 31.5, 63.5$ respectively.} \label{f7}
\end{figure}

In Fig. \ref{f7}, the Binder cumulant is plotted for the $3$D
Ising model starting from the semi-ordered state. For a small $s$,
one observes the power-law behavior at bulk, $U(t,x) \sim t^{(d -
1)/z}$. From the slope of the curve, one obtains $0.995(12)$, and
then estimates $z=2.01(2)$, consistent with $z = 2.04(1)$ from the
literature \cite{jas99}. For a large $s$, e.g., $x = 0.5$ and
$t>10$, one estimates $(d - 1 + d_0)/z=1.963(10)$ from the
power-law behavior in Eq.~(\ref{equ165}), then derives $ d_0 =
2.01(2)$. Again it is close to $2$. To further verify the scaling
form in Eq.~(\ref{equ70}), we fix $x' =1.5$, and change the time
scale $t$ of another $x$ to $(x'/x)^z\ t$, and the scale of
$U(t,x)$ to $(x'/x)^{d-1}U(t,x)$. As shown in Fig.~\ref{f6} and
\ref{f7}, all curves of different $x$ nicely collapse to the curve
of $x'=1.5$. This fully confirms the scaling form in
Eq.~(\ref{equ70}).

\begin{figure}[p]
\epsfysize=6.cm \epsfclipoff \fboxsep=0pt
\setlength{\unitlength}{1.cm}
\begin{picture}(6,6)(0,0)
\put(-1.0,0.3){{\epsffile{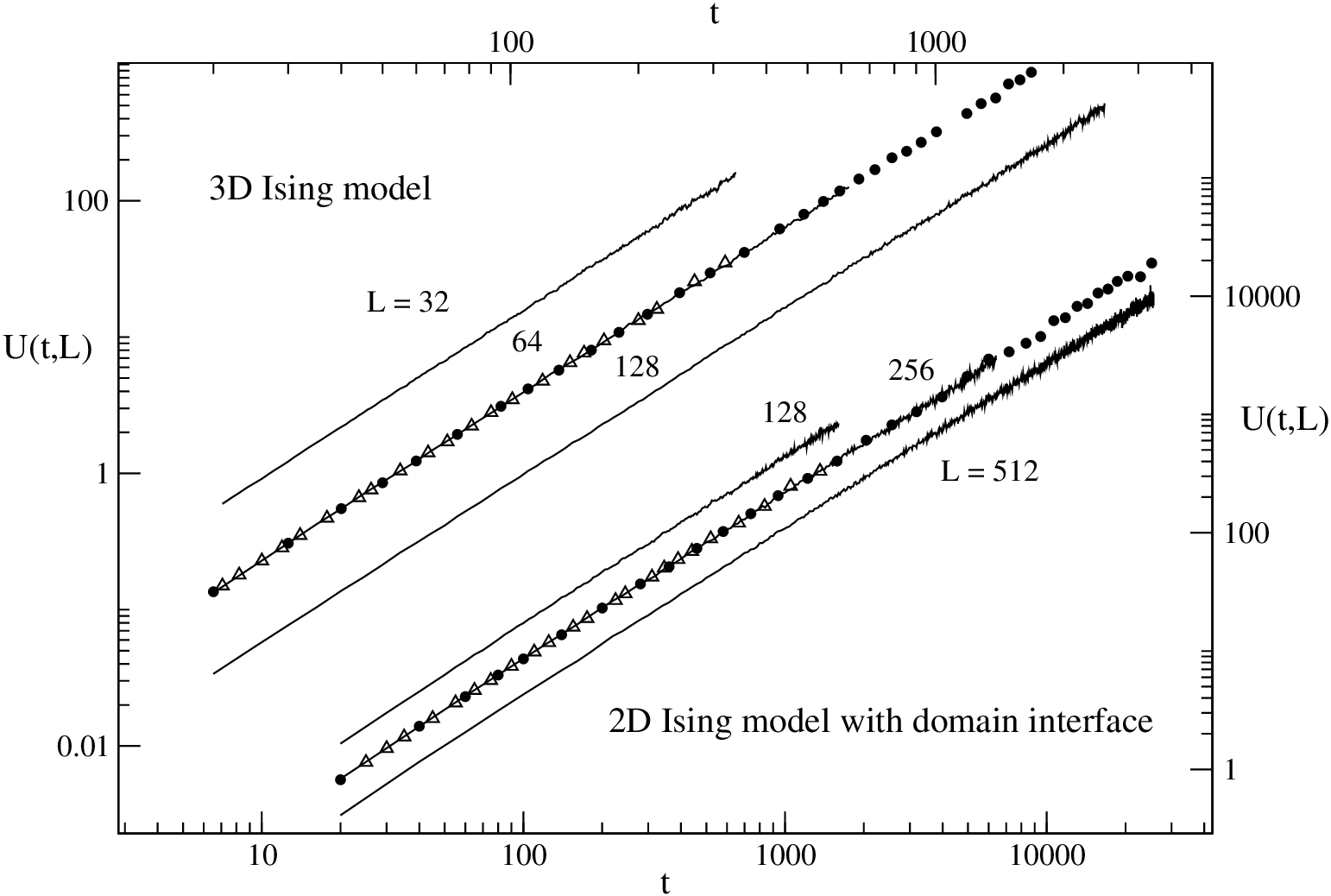}}}
\end{picture}
\caption{The Binder cumulant at $x=0.5$ for the Ising model with
the domain interface is plotted with solid lines on a double-log
scale. The lower three solid lines are for the $2$D Ising model
with the lattice size $L = 128$, $256$ and $512$. The $x-$ and
$y-$axis are on the bottom and right sides. According to
Eq.~(\ref{equ70}), the curves collapse to the curve at $L' = 256$.
Circles and triangles correspond to $L = 512$ and $128$
respectively. The upper three solid lines are for the $3$D Ising
model with $L = 32$, $64$ and $128$. The $x-$ and $y-$axis are on
the top and left sides. Data collapse is observed. Circles and
triangles correspond to $ L = 128$ and $32$ respectively.}
\label{f9}
\end{figure}

To reveal the lattice-size dependence of the Binder cumulant in
Eq.~(\ref{equ70}), we fix $x = 0.5$, and plot $U(t,L) \equiv
U(t,x=0.5)$ as a function of $t$ for different $L$ in
Fig.~\ref{f9}. Obviously, all curves of different $L$ and in two-
and three-dimensions are parallel each other, and exhibit the
power-law behavior in Eq.~(\ref{equ165}). We then fix a lattice
size $L'$, and change the scale of $U(t,L)$ of another $L$ to
$U(t,L)(L'/L)^{-(d - 1)}$. Data collapse is clearly observed for
both the $2$D and $3$D Ising models.

\begin{figure}[p]
\epsfysize=6.cm \epsfclipoff \fboxsep=0pt
\setlength{\unitlength}{1.cm}
\begin{picture}(6,6)(0,0)
\put(-1.0,0.3){{\epsffile{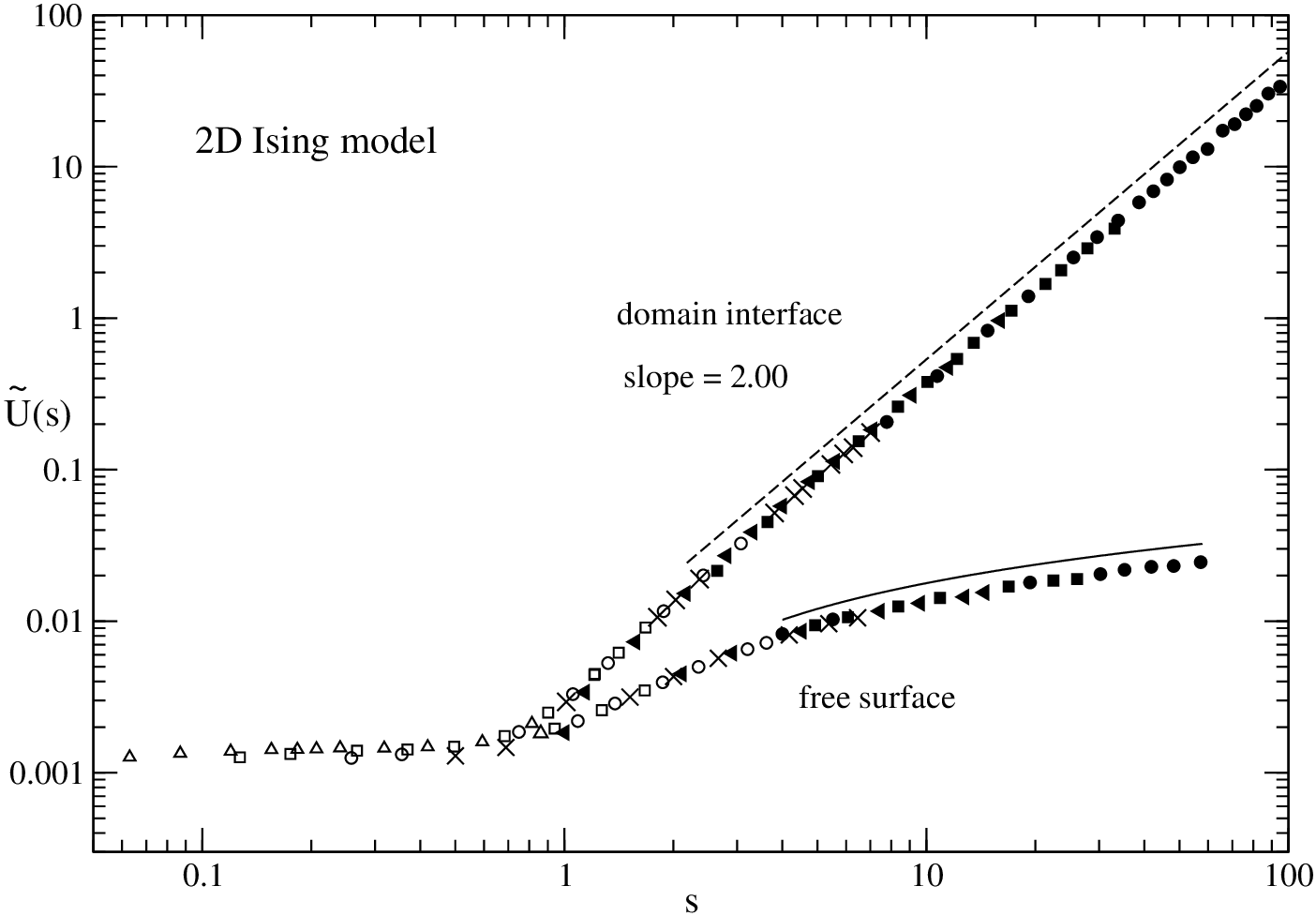}}}
\end{picture}
\caption{The scaling function $\widetilde{U}(s)$ with
$s=t^{1/z}/x$ is plotted on a double-log scale, for the $2$D Ising
model with the domain interface and free surface. Data collapse
for different $x$ is observed. The dashed line shows the power law
fit, and the solid line indicates the logarithmic fit. }
\label{f8}
\end{figure}

In Fig.~\ref{f8}, the scaling function $\widetilde{U}(s)$ with
$s=t^{1/z}/x$ is plotted for the $2$D Ising model with the domain
interface and free surface. Data of different $x$ collapse clearly
onto their master curves. For the domain interface, the asymptotic
behavior of $\widetilde{U}(s)$ in Eq.~(\ref{equ160}) is exposed.
The exponent $d_0$ is measured to be $2.00(2)$, the same as that
extracted from the single curve of $x=0.5$ in Fig.~\ref{f6}. For
the free surface, $\widetilde{U}(s) \to const$ is also observed in
the limit $s \to 0$. In the large $s$ regime, however, it does not
exhibit a power-law behavior. Instead, it increases by a
logarithmic law, $\widetilde{U}(s) = a_0 + a_1\ln(s)$. In other
words, the exponent $d_0$ of the free surface is effectively $0$
but with a logarithmic correction. This result indicates that the
spatial fluctuation of the domain interface grows in time much
faster than that of the free surface.

In three dimensions, $d - 1 - 2\beta_1/\nu$ of the free surface is
negative. Starting from an ordered state, the susceptibility
decays in time. Therefore, one suffers from large fluctuations,
and it is difficult to address the dynamic behavior of the
susceptibility or Binder cumulant. For the disordered surface, the
situation is even more complicated. Since our paper is already
lengthy, we will not go into the details here.

\subsection{Height function and roughness function}

In the preceding two subsections, we have analyzed the temporal
and spatial structures of the magnetization $M(t,x)$ and Binder
cumulant $U(t,x)$. Up to now, however, we have not yet touched how
the interface grows and fluctuates in the $x$ direction. For this
purpose, we have introduced the height function $h(t)$ and the
roughness function $\omega^2(t)$ in Eqs.~(\ref{equ40}) and
(\ref{equ50}). For a standard growing interface, the time
evolution of the height function $h(t)$ may be not so important,
but the roughness function $\omega^2(t)$ exhibits a power-law
behavior governed by the roughness exponent.

\begin{figure}[p]
\epsfysize=6.cm \epsfclipoff \fboxsep=0pt
\setlength{\unitlength}{1.cm}
\begin{picture}(6,6)(0,0)
\put(-1.0,0.3){{\epsffile{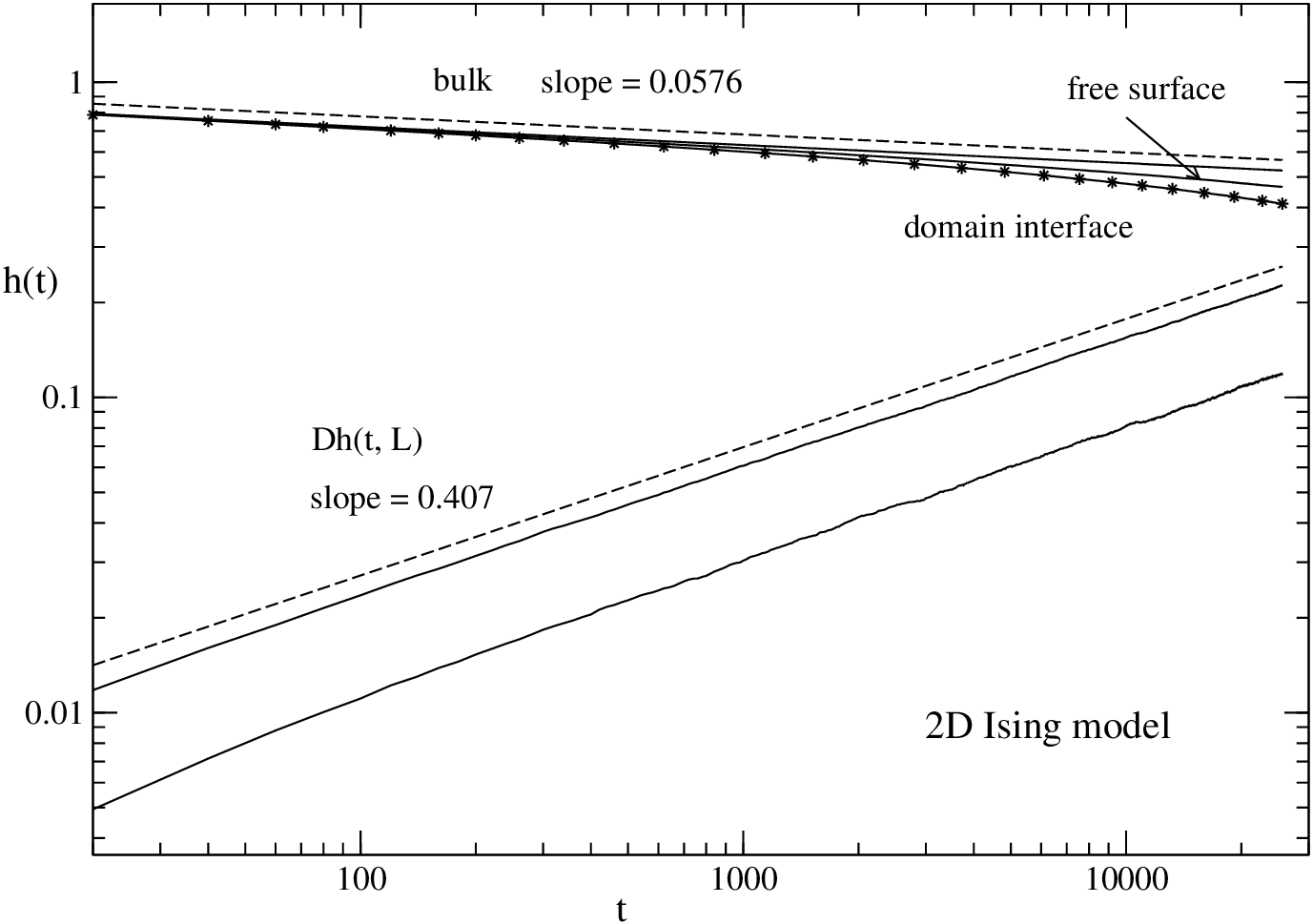}}}
\end{picture}
\caption{The height function $h(t)$ of the $2$D Ising model is
plotted on a double-log scale. The three upper solid lines are for
the domain interface, free surface and bulk (from below). The
lower solid lines are the pure height function $Dh(t,L)$ of the
free surface and domain interface (from below). Dashed lines show
the power-law fits, and stars are from a double power-law fit.}
\label{f13}
\end{figure}

In Fig.~\ref{f13}, the height function $h(t)$ is plotted for the
$2$D Ising model. One finds a power-law behavior for the dynamic
relaxation of the bulk, i.e., with periodic boundary conditions in
all directions. The slope of the curve is $0.0576(3)$, consistent
with $\beta / \nu z = 0.0579(3)$ from the literature \cite{zhe98}.
For the domain interface and free surface, the height function
decreases faster than a power law. Actually, the curves can be
fitted by a double power law, e.g.,
$h(t)=c_0t^{\alpha_0}-c_1t^{\alpha_1}$. Although this
four-parameter fit could not produce very accurate values of
$\alpha_0$ and $\alpha_1$, it leads us to introduce the pure
height function $Dh(t,L)$ in Eq.~(\ref{equ85}). The conjecture is
that the term $c_1t^{\alpha_1}$ represents the pure interface, and
$c_0t^{\alpha_0}$ is the magnetization of the bulk. In
Fig.~\ref{f13}, we do observe a power-law behavior for the pure
height function $Dh(t,L)$. The slope of the curves is estimated to
be $0.407(2)$. In comparison with that for a standard growing
interface, this power-law behavior is special for the domain
interface.

\begin{figure}[p]
\epsfysize=6.cm \epsfclipoff \fboxsep=0pt
\setlength{\unitlength}{1.cm}
\begin{picture}(6,6)(0,0)
\put(-1.0,0.3){{\epsffile{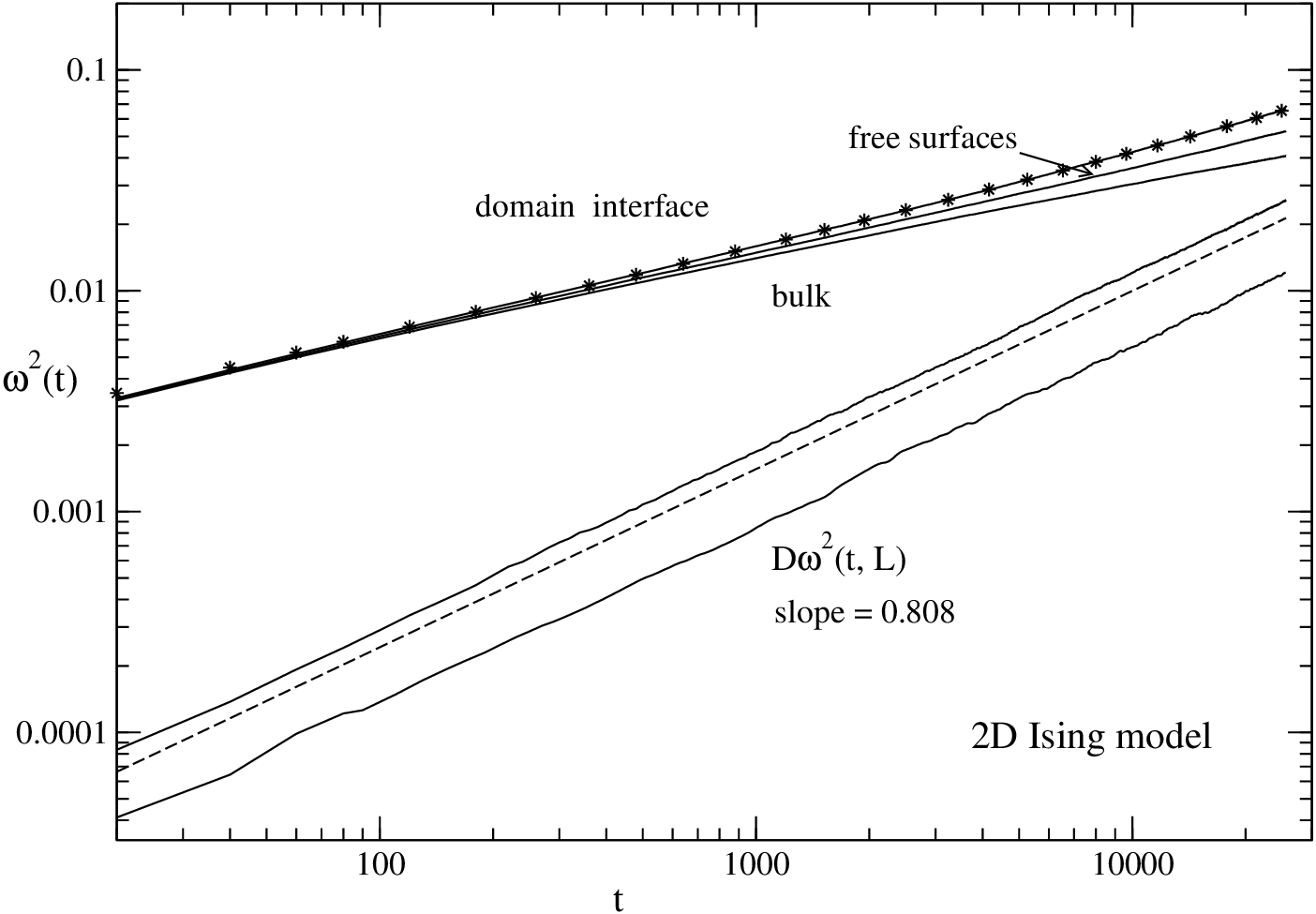}}}
\end{picture}
\caption{The roughness function $\omega^2(t)$ of the $2$D Ising
model is plotted on a double-log scale. The three upper solid
lines are for the domain interface, free surface and bulk (from
above). The lower solid lines are the pure roughness function
$D\omega^2(t,L)$ for the domain interface and free surface (from
above). Dashed lines show the power-law fits, and stars are from a
double power-law fit.} \label{f11}
\end{figure}

In Fig. \ref{f11}, the roughness function is plotted for the $2$D
Ising model. In the case of the bulk, $\omega^2(t)$ is expected to
obey a power law, although there exist corrections to scaling.
Anyway, one may roughly estimate the exponent $(1-2\beta / \nu)/z$
to be $0.345(14)$, consistent with $z = 2.16(1)$. Due to
corrections to scaling, the dynamic behavior of $\omega^2(t)$
looks unclear for the domain interface and free surface. However,
the pure roughness function $D\omega^2(t,L)$ in Eq.~(\ref{equ87})
obviously obeys a power-law behavior for both the domain interface
and free surface. It seems that $\omega^2(t)$ and $\omega_b^2(t)$
have the same correction to scaling. Neglecting the corrections to
scaling, one may assume
\begin{equation}
   \widetilde{\omega^2}(u) =  \{
   \begin{array}{lll}
    \mbox{c}    & \quad &  \mbox{without interface} \\
   c+u^{d_{\omega}}  & \quad &  \mbox{with interface}
   \end{array}.
   \label{equ195}
\end{equation}
Then one derives
\begin{equation}
D\omega^2(t,L) =  \frac{t^{(1 - 2\beta / \nu +
d_{\omega})/z}}{L^{1 + d_{\omega}}}. \label{equ200}
\end{equation}
In Fig. \ref{f11}, the curves of $D\omega^2(t,L)$ for the domain
interface and free surface are parallel each other. From the
slopes of the curves one measures $(1 - 2\beta / \nu +
d_{\omega})/z=0.808(2)$. Then one calculates the exponent
$d_{\omega} = 0.995(4)$. The fluctuations of the domain interface
and free surface grow faster in time than that of the bulk. More
importantly, the exponent $(1 - 2\beta / \nu + d_{\omega})/z$ is
independent of the interface exponent or surface exponent
$\beta_1$.

\begin{figure}[p]
\epsfysize=6.cm \epsfclipoff \fboxsep=0pt
\setlength{\unitlength}{1.cm}
\begin{picture}(6,6)(0,0)
\put(-1.0,0.3){{\epsffile{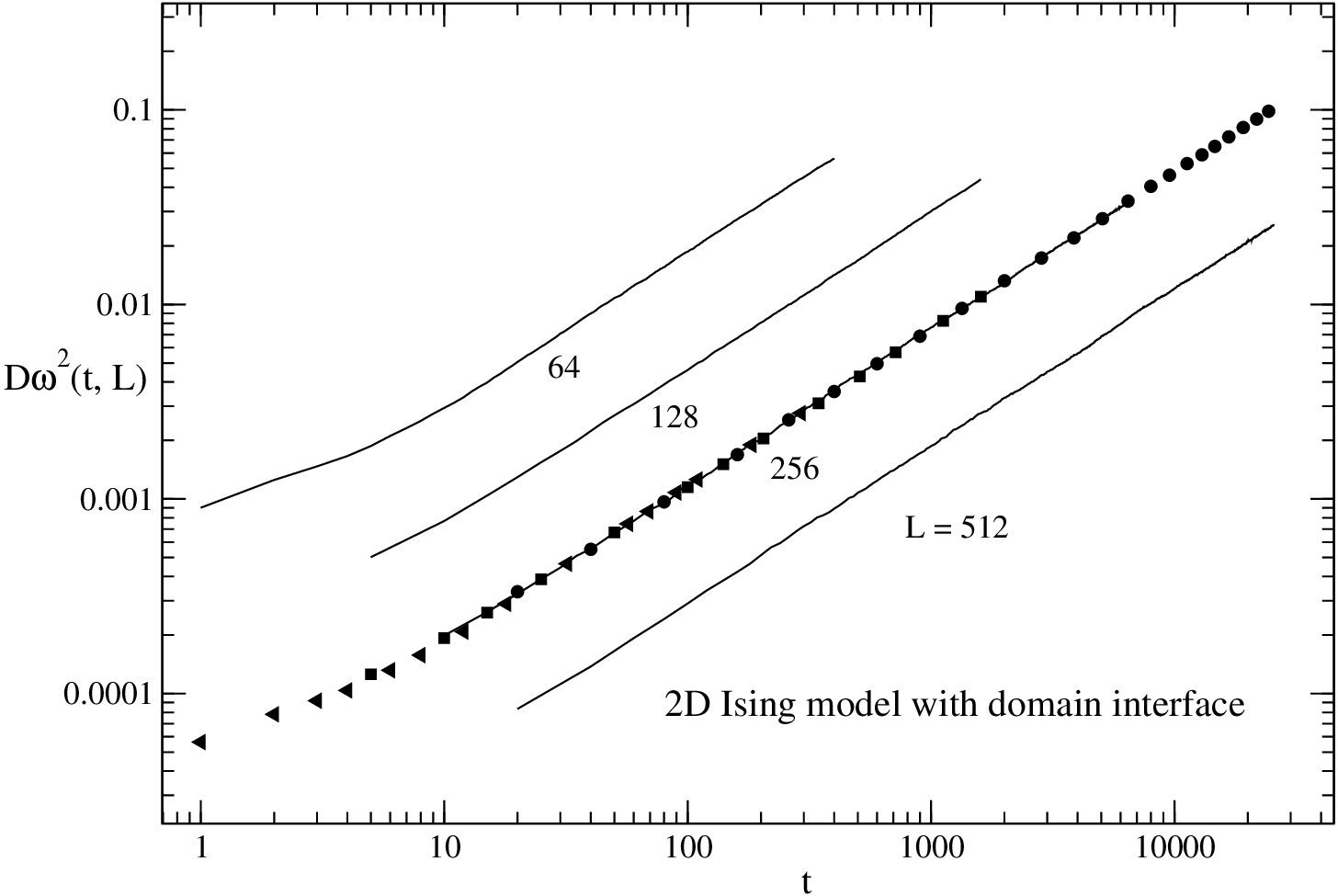}}}
\end{picture}
\caption{The pure roughness function $D\omega^2(t,L)$ of the
domain interface is plotted for the $2$D Ising model with solid
lines on a double-log scale, for $ L = 512$, $256$, $128$ and $64$
(from below). According to Eq.~(\ref{equ200}), data collapse is
observed at the curve of $L' = 256$. Solid circles, solid squares,
and solid triangles correspond to $L = 512$, $128$ and $64$
respectively.} \label{f12}
\end{figure}

One may also verify the lattice-size dependence, $D\omega^2(t,L)
\sim 1/L^{1 + d_{\omega}}$ in Eq.~(\ref{equ200}). In
Fig.~\ref{f12}, $D\omega^2(t,L)$ is plotted for different lattice
sizes. Obviously, all the curves are parallel each other. Then we
fix a lattice size, e.g., $L'=256$, and change the scale of
$D\omega^2(t,L)$ of another $L$ to
$D\omega^2(t,L)(L'/L)^{-(1+d_{\omega})}$. Taking $d_{\omega}=1$ as
input, data collapse is clearly observed.

For a standard growing interface, the roughness function grows by
$\omega^2(t) \sim t^{2\alpha/z}$, and $\alpha$ is the so-called
roughness exponent. According to Eq.~(\ref{equ200}), the roughness
exponent for the domain interface and free surface is $\alpha = (1
- 2 \beta / \nu + d_{\omega}) /2 =0.872(8)$. On the other hand,
from the dimension counting one may expect $Dh(t,L) \sim
t^{\alpha/z}$ for the pure height function. From the measurement
$\alpha/z=0.407(2)$ in Fig.~\ref{f13}, one calculates $\alpha
=0.879(6)$. These two measurements of the exponent $\alpha$ are in
good agreement with each other. In the scaling analysis of
$D\omega^2(t,L)$, $- \beta / \nu$ represents the scaling dimension
of the magnetization. One may remove it, e.g., by dividing
$D\omega^2(t,L)$ by $M(t)^2$ of the bulk. Then the real roughness
exponent is $(1  + d_{\omega}) /2$. Since $d_{\omega} = 1$, the
roughness exponent is just $1$. This conclusion holds also for the
Ising model in three dimensions. But the measurements of the
exponents are somewhat complicated in this case, for $1 - 2 \beta
/ \nu=-0.038$ is around zero.

\section{Conclusion}

In summary, we have investigated the non-equilibrium critical
dynamics with a domain interface, a free surface and a disordered
surface, taking the two- and three-dimensional Ising models as
examples. The dynamic scaling behavior is revealed, and a dynamic
roughening process is observed. Critical exponents characterizing
the magnetization, Binder cumulant, height function and roughness
function are extracted, and the results are summarized in Table
\ref{t1}.

i) For the domain interface, $\beta_0 / \nu$ for the magnetization
in Eq.~(\ref{equ110}) takes values close to $1$ for both the two-
and three-dimensional Ising models. It indicates that the
magnetization $M(t,x)$ is an analytic function of $x$. Especially,
$M(t,x)$ {\it inside} the domain interface decays much faster in
time than that {\it at bulk}, for the interface exponent $\beta_1=
\beta+\beta_0$ is much bigger than the bulk exponent $\beta$. For
the free surface, the values of $\beta_1\equiv \beta_s$ are in
agreement with the measurements in equilibrium. For the disordered
surface, $\beta_1\equiv \beta^*_s$ takes the same value as that of
the free surface, but with a logarithmic correction to scaling in
two dimensions.

ii) For the domain interface, the exponent $d_0$ for the Binder
cumulant in Eq.~(\ref{equ160}) takes values close to $2$ in two
and three dimensions. For the free surface, $d_0\equiv d_s$ is
effectively $0$ in two dimensions, but with a logarithmic
correction to scaling. These results indicate that the fluctuation
in the $y$ direction inside the domain interface is stronger than
that around the free surface. In fact, one can derive from
Eqs.~(\ref{equ110}) and (\ref{equ160}) that inside the domain
interface, the susceptibility behaves as $M^{(2)}(t,x)-M(t,x)^2
\sim t^{(d - 1 -2\beta/\nu)/z}$, the same as that at bulk. Around
the free surface, the susceptibility is $M(t,x)^{(2)}-M(t,x)^2
\sim t^{(d - 1 -2\beta_1/\nu)/z}$, different from that at bulk.

iii) For both the domain interface and free surface, the roughness
function in Eq.~(\ref{equ80}) does not obey a power law, for it
includes the fluctuation of the bulk and domain interface. After
subtracting the contribution of the bulk, the pure roughness
function in Eq.~(\ref{equ87}) does exhibit a power-law behavior in
Eq.~(\ref{equ200}), and the roughness exponent is identified to be
$\alpha=(1 + d_{\omega})/2$. Interestingly, the exponent
$d_{\omega}$ takes values close to $1$ for both the domain
interface and free surface, and also independently of the spatial
dimension. In other words, the fluctuation of the interface in the
$x$ direction is independent of the exponent $\beta_1$.

Theoretically, above results need further understanding. For
example, it is a challenge to derive the dynamic scaling forms
with renormalization group methods. It is also important to
investigate how the quenched disorder may affect the domain-wall
motion at the critical temperature. The techniques used in this
paper may be also applied to similar dynamic systems.

{\bf Acknowledgements:} This work was supported in part by NNSF
(China) under grant No. 10325520.


\begin{table}[h]\centering
\begin{tabular}[t]{c|l|l|l}
\hline  &    &  $2$D Ising &  $3$D Ising\\
\hline
          $M(t)$    & $\beta_0 / \nu$    & 0.998(5)         &    1.001(6)         \\
                    & $\beta_1$          & 1.123(5)         &    0.958(6)         \\
\hline              & $\beta_1\equiv \beta_s $         & 0.499(4)         &    0.801(6)     \\

                    & $\beta_1\equiv \beta^*_{s} $     & 0.499            &    0.812(4)     \\
\hline    $U(t)$    & $d_0$              &  2.00(2)         &   2.01(2)       \\
                    & $d_s$              &  0               &        \\
\hline
               $\omega^2(t)$  & $d_{\omega}$     & 0.995(4)         &   $1$   \\
\hline  \hline
           & $\beta_s$   & 1/2\quad\,  \cite{ple04a}       &    0.795(10) \cite{bel06}            \\
           & $\beta$  &  1/8         &  0.327(1)\,\,\,\,\cite{jas99}   \\
           &  $\nu$   &  1           &  0.630(2)\,\,\, \cite{jas99}  \\
           &  $z$     & 2.16(1) \cite{zhe98}     &  2.04(1)\,\,\,\,\,  \cite{jas99}   \\
\hline
\end{tabular}
\caption{In the upper sector, the exponents $\beta_0 / \nu$,
$\beta_1$ and $d_0$ are for the domain interface, $\beta_s$ and
$d_s$ are for the free surface, and $\beta^*_{s} $ is for the
disordered surface. The exponent $d_{\omega}$ is for both the
domain interface and free surface. In the measurements of
$\beta^*_{s}$ and $d_s$ for the $2$D Ising model, logarithmic
corrections to scaling are taken into account. In the lower
sector, the static exponents $\beta$ and $\nu$, the dynamic
exponent $z$, and the surface exponent $\beta_s$ are taken from
the literatures.} \label{t1}
\end{table}

\end{document}